\documentclass[10pt,letterpaper]{article}
\usepackage{opex3}
\usepackage{amssymb,amsmath}
\usepackage{dsfont}

\begin{document}
\title{Experimental demonstration of decoherence suppression via \\quantum measurement reversal}

\author{Jong-Chan Lee, Youn-Chang Jeong, Yong-Su Kim, \\and Yoon-Ho Kim$^*$}
\address{Department of Physics, Pohang University of Science and Technology (POSTECH), \\Pohang, 790-784, Korea}
\email{$^*$yoonho72@gmail.com}\homepage{http://qopt.postech.ac.kr}

\date{\today}

\begin{abstract}
Taming decoherence is essential in realizing quantum computation and quantum communication. Here we experimentally demonstrate that decoherence due to amplitude damping can be suppressed by exploiting quantum measurement reversal in which a weak measurement and the reversing measurement are introduced before and after the decoherence channel, respectively. We have also investigated the trade-off relation between the degree of decoherence suppression and the channel transmittance.
\end{abstract}

\ocis{(270.5585) Quantum information and processing; (270.250) Fluctuations, relaxations, and noise; (270.5565) Quantum communications;}



\section{Introduction}

The benefits of quantum computation and quantum communication surpassing the limit of classical information technology rely on coherent evolution and transmission of quantum states (i.e., qubits or qudits). Decoherence refers to loss of quantum coherence due to, for example, depolarization, amplitude damping, and/or phase damping, affecting both quantum superposition of single-quantum states and multipartite entanglement \cite{Nielsen,Aolita,Almeida}. Suppressing the decoherence in quantum channels, thus, is one of the biggest challenges in quantum information.

Decoherence suppression may be accomplished in a number of ways, including quantum error correction \cite{Shor,Steane}, decoherence-free subspace \cite{Lidar,Kwiat}, and dynamical decoupling \cite{Viola,West}. In the quantum error correction approach to decoherence suppression, the error correction operation transforms entanglement between the environment and the system into entanglement between the environment and the ancilla \cite{Shor,Steane}. In the decoherence-free subspace approach, quantum information is encoded in a particular quantum state that does not experience a specific type of decoherence \cite{Lidar,Kwiat}. Note that both the quantum error correction and the  decoherence-free subspace schemes utilize the Hilbert space dimension larger than that of the system alone, requiring more resources for implementation. The dynamical decoupling method achieves decoherence suppression without increasing the Hilbert space dimension, but it cannot be applied to Markovian processes (e.g., decoherence due to amplitude damping) where the characteristic time for decoherence is shorter than the practicable control time \cite{Viola,West}.

In this paper, we report an experimental demonstration of decoherence suppression via quantum measurement reversal recently proposed in Ref.~\cite{Korotkov}. Decoherence due to amplitude damping is shown to be effectively suppressed by introducing a weak quantum measurement that partially collapses an initial quantum state and the reversing measurement that restores the original quantum state before and after the decoherence channel, respectively. The decoherence suppression scheme works without introducing a larger Hilbert space but at the reduced transmittance of the channel.

\section{Theory}

Decoherence due to amplitude damping can be understood as a process in which the excitation of the system qubit is transferred to that of the environment qubit \cite{Nielsen},
\begin{equation}
|1\rangle_S\otimes|0\rangle_E\rightarrow \sqrt{1-D}|1\rangle_S
\otimes|0\rangle_E+\sqrt{D}|0\rangle_S\otimes|1\rangle_E
,\label{eq1}
\end{equation}
where the subscripts $S$ and $E$ denote the system and the environment, respectively, and $D$ is the magnitude of the decoherence. The computational basis state $|0\rangle_S$ is not affected in this type of decoherence and the effect of decoherence is to probabilistically and incoherently collapse an initial quantum state toward the state $|0\rangle_S$.


The decoherence suppression scheme makes use of the asymmetric nature of amplitude damping decoherence \cite{Nielsen,Korotkov} and the fact that weak quantum measurement can be reversed \cite{Koashi, Kim, Korotkov2006, Katz}. The initial quantum state, before encountering the decoherence channel, undergoes the weak quantum measurement to cause partial collapse of the initial state toward the $|0\rangle_S$ state which experiences no decoherence. After the decoherence channel, the reversing measurement is applied to probabilistically reject the $|0\rangle_S$ state that includes the effect of state collapse due to decoherence. By applying two non-unitary operations before and after the amplitude damping channel, the decoherence in the channel can be efficiently suppressed whereas the channel transmittance is reduced. The channel loss originates solely from the non-unitary nature of the weak and the reversing measurement.


Let us now describe in detail the decoherence suppression scheme via quantum measurement reversal proposed in Ref.~\cite{Korotkov}. As mentioned above, decoherence under consideration is of the amplitude damping type shown in Eq.~(\ref{eq1}). Initially, the system qubit is prepared in $|\psi\rangle_S=\alpha|0\rangle_S+\beta|1\rangle_S$ with $|\alpha|^2+|\beta|^2=1$ and the environment qubit is initialized to $|0\rangle_E$. First, the weak quantum measurement (or partial
collapse measurement) with strength $p$ is applied to the system qubit \cite{Kim, Korotkov2006, Katz}, causing the initial state partially collapsed to
\begin{equation}
|\psi_1\rangle = \left(\frac{\alpha|0\rangle+\beta\sqrt{1-p}|1\rangle}{\sqrt{|\alpha|^2+
|\beta|^2(1-p)}}\right)_S\otimes|0\rangle_E.\label{eq2}
\end{equation}
Here we only consider the measurement outcome that is reversible and the probability for this event is $P_1=|\alpha|^2+|\beta|^2(1-p)$. The irreversible outcome (i.e., due to projection measurement) is discarded. Note that, by causing partial collapse of the system qubit, the initial state has moved closer to the $|0\rangle_S$ state that does not experience amplitude damping decoherence. After the amplitude damping decoherence channel with the magnitude of decoherence $D$, the system-environment qubit state changes to,
\begin{equation}
|\psi_2\rangle = \frac{1}{\sqrt{P_1}}\left[\big(\alpha|0\rangle+\beta\sqrt{1-p}\sqrt{1-D}
|1\rangle\big)_S\otimes
|0\rangle_E+\beta_D|0\rangle_S\otimes|1\rangle_E\right],\label{eq3}
\end{equation}
where $|\beta_D|^2=|\beta|^2(1-p)D$ is the probability that the system qubit would experience decoherence. Finally, the reversing measurement with strength $p_r$ is applied, causing the following state change:
\begin{equation}
|\psi_3\rangle=\frac{1}{\sqrt{T}}\left[\big(\alpha\sqrt{1-p_r}|0\rangle+\beta\sqrt{1-p}\sqrt{1-D}
|1\rangle\big)_S\otimes
|0\rangle_E+\beta_D\sqrt{1-p_r}|0\rangle_S\otimes|1\rangle_E\right].\label{eq4}
\end{equation}
The transmittance, $T=|\alpha|^2(1-p_r)+|\beta|^2(1-p)(1-D)+|\beta_D|^2(1-p_r)$, is defined as the probability that the initial qubit state $|\psi\rangle_S\otimes|0\rangle_E$ is non-unitarily transformed to the final state $|\psi_3\rangle$ via the sequence of weak measurement, decoherence, and reversing measurement without being irreversibly collapsed. We define the channel transmittance as the transmittance averaged over the Bloch sphere, $T_{ch}=\left\langle T\right\rangle_{avg}=\frac{1}{2}(1-p_r)+\frac{1}{2}(1-p)(1-p_rD).$ The channel transmittance is reduced as the weak measurement $p$ and the reversing measurement strength $p_r$ are increased.

If we have \textit{a priori} knowledge about the magnitude of decoherence $D$, we can perform the optimal reversing operation with $p_r=p+D(1-p)$ \cite{Korotkov}. After applying the optimal reversing measurement, the system qubit state is calculated to be 
\begin{eqnarray}
\rho_S^f=\frac{P_R|\psi\rangle_S\langle\psi|+P_D|0\rangle_S\langle0|}{P_R+P_D},
\label{eq5}
\end{eqnarray}
where $P_R=(1-p)(1-D)$ and $P_D=|\beta|^2(1-p)^2D(1-D)$. Note that $P_D/P_R=|\beta|^2(1-p)D$ is a monotonically decreasing function of $p$. Hence, if the initial weak measurement strength $p$ is increased to 1, the final state $\rho_S^f$ can be made arbitrarily close to the initial state $|\psi\rangle_S$ which means that decoherence is fully suppressed. The trade-off is that the channel transmittance $T_{ch}$ will decrease as stronger decoherence suppression is applied.

\section{Experimental setup}


\begin{figure}[t]
\centering
\includegraphics[width=3.7in]{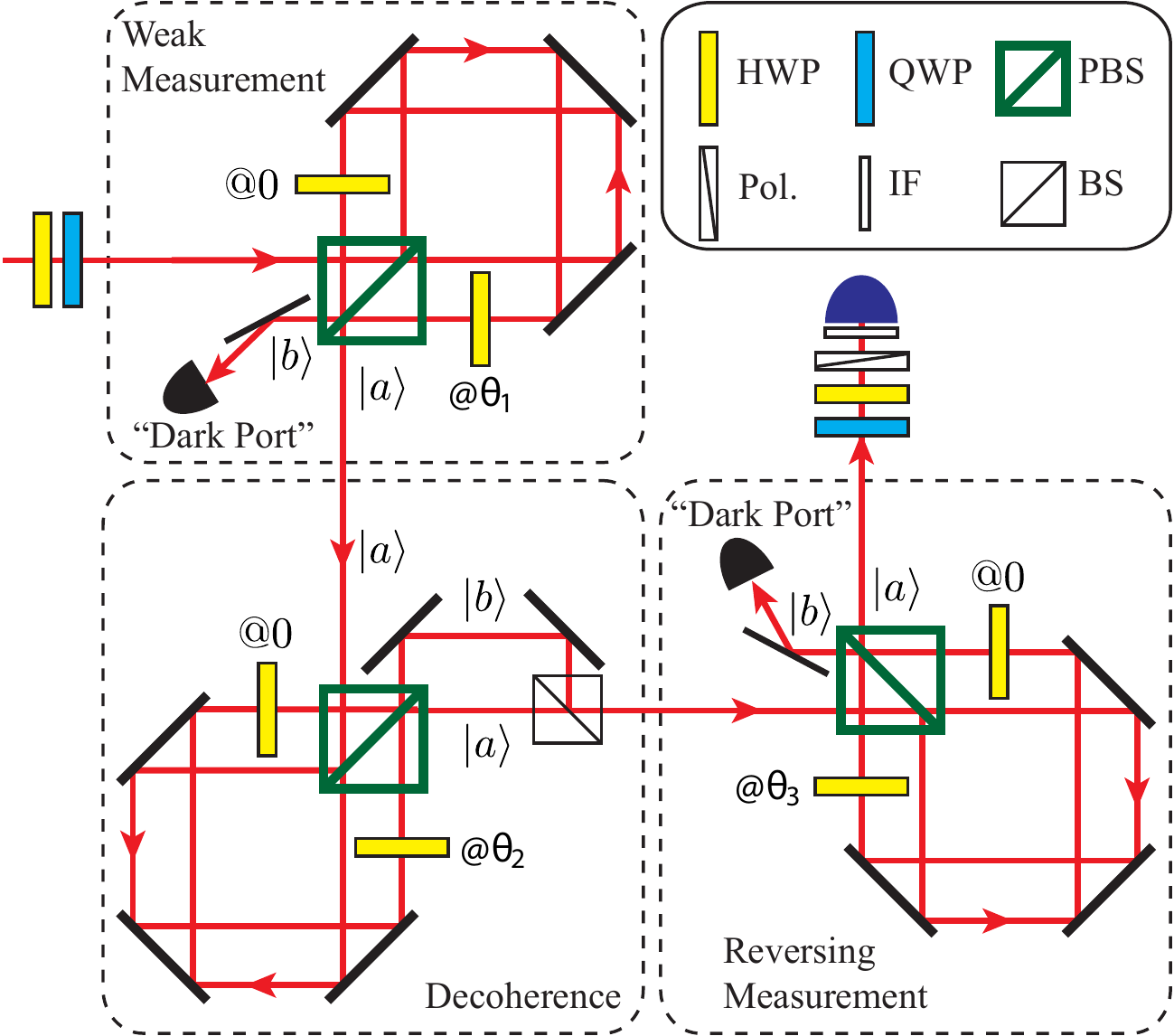}
\caption{The system qubit is the single-photon polarization qubit, prepared using a half-wave plate (HWP) and a quarter-wave plate (QWP), and the environment qubit is the path qubit with the basis states $|a\rangle$ and $|b\rangle$. The Sagnac-type interferometers (SI) implement the weak measurement and the reversing measurement.
The amplitude damping decoherence shown in Eq.~(\ref{eq1}) is realized using a SI with an additional beam splitter (BS) which implements tracing out of the environment qubit. Quantum state tomography is performed on the output polarization qubit.}
\label{scheme}
\end{figure}

In the following, we report photonic realization of decoherence suppression via quantum  measurement reversal with the single-photon polarization qubit as the system qubit and the single-photon path qubit as the environment qubit.
The experimental schematic is shown in Fig.~\ref{scheme}. Let us first describe the single-photon source used in this experiment (not shown in Fig.~\ref{scheme}). A 810 nm photon pair is generated from the spontaneous parametric down-conversion (SPDC) process in a 10 mm long type-II PPKTP crystal pumped by a 405 nm diode laser \cite{hong,ppktp}. The orthogonally polarized signal-idler photon pair is spatially split by a polarizing beam splitter. The detection of the idler photon heralds preparation of the single-photon state for the signal photon. The heralded single-photon state is then polarization encoded with a half-wave plate (HWP) and a quarter-wave plate (QWP) to form the single-photon polarization qubit $|\psi\rangle_S = \alpha |H\rangle + \beta |V\rangle$. Note that $|H\rangle \equiv |0\rangle_S$ and $|V\rangle \equiv |1\rangle_S$.

The weak measurement (i.e., partial collapse measurement) and the reversing measurement on the system qubit, i.e., the single-photon polarization qubit, can be accomplished by preferentially and partially reducing one of the two amplitudes of the polarization basis states \cite{Kim}. To vary the measurement strength $p$ and  $p_r$ continuously, the weak measurement and the reversing measurement are implemented with the Sagnac-type interferometer (SI)
consisting of a polarizing beam splitter (PBS), three mirrors and two half-wave plates (HWPs), see Fig.~\ref{scheme}. The HWPs labelled with `@$\theta_1$' and `@$\theta_3$' can be rotated to certain angles to vary $p$ and $p_r$ values respectively and the HWPs labelled with `@0' is used to keep the path lengths identical. The SI-based weak measurement has two outcomes
(or output modes):  For the single-photon found in the $|a\rangle$ output mode, the input qubit state  $|\psi\rangle_S$ changes to $\alpha|0\rangle_S+\beta\sqrt{1-p}|1\rangle_S$ with  $\sqrt{p}=\sin2\theta_1$ and this case corresponds to the weak measurement on the polarization qubit which may be reversed. In the output mode $|b\rangle$ (labelled as ``dark port''), the qubit state is found to be $\beta\sqrt{p}|0\rangle_S$ and this corresponds to projection measurement which is irreversible. The SI-based reversing measurement has the same effect on the input qubit but with $|0\rangle$ and $|1\rangle$ interchanged. Note again that the ``dark port'' output of the SI corresponds to irreversible collapse of the system qubit state so any events at the dark port should be discarded. However, since we observe the coincidence between the heralding detector and the signal detector, there is no need for actually monitoring the dark port, see Ref.~\cite{Kim}.

The amplitude damping decoherence effect shown in Eq.~(\ref{eq1}) is realized with a SI with an additional beamsplitter (BS) which implements tracing out of the environment qubit, see Fig.~\ref{eq1}. The environment qubit is the path qubit of the single-photon state $|a\rangle \equiv |0\rangle_E$ and $|b\rangle \equiv |1\rangle_E$ \cite{Almeida}. For the single-photon state $|V\rangle$ entering the SI via the $|a\rangle$ mode, the state at the output is
\begin{equation}
|V\rangle\otimes|a\rangle\rightarrow \cos2\theta_2|V\rangle
\otimes|a\rangle+\sin2\theta_2|H\rangle\otimes|b\rangle,\label{eq6}
\end{equation}
where $\theta_2$ is the HWP angle. In contrast, the state $|H\rangle$ is unchanged, i.e. $|H\rangle\otimes|a\rangle\rightarrow|H\rangle\otimes|a\rangle$. Note that the above expression is identical to Eq.~(\ref{eq1}) if $\sqrt{D}=\sin 2\theta_2$. Since we are interested in the system qubit only, the environment qubit is traced out by incoherently mixing $|a\rangle$ and $|b\rangle$ at the BS. Note that the path length difference between $|a\rangle$ and $|b\rangle$ is much larger than the coherence length of the down-conversion photons ($\sim70 \mu$m).

 \begin{figure}[b]
\centering
\includegraphics[width=3.3in]{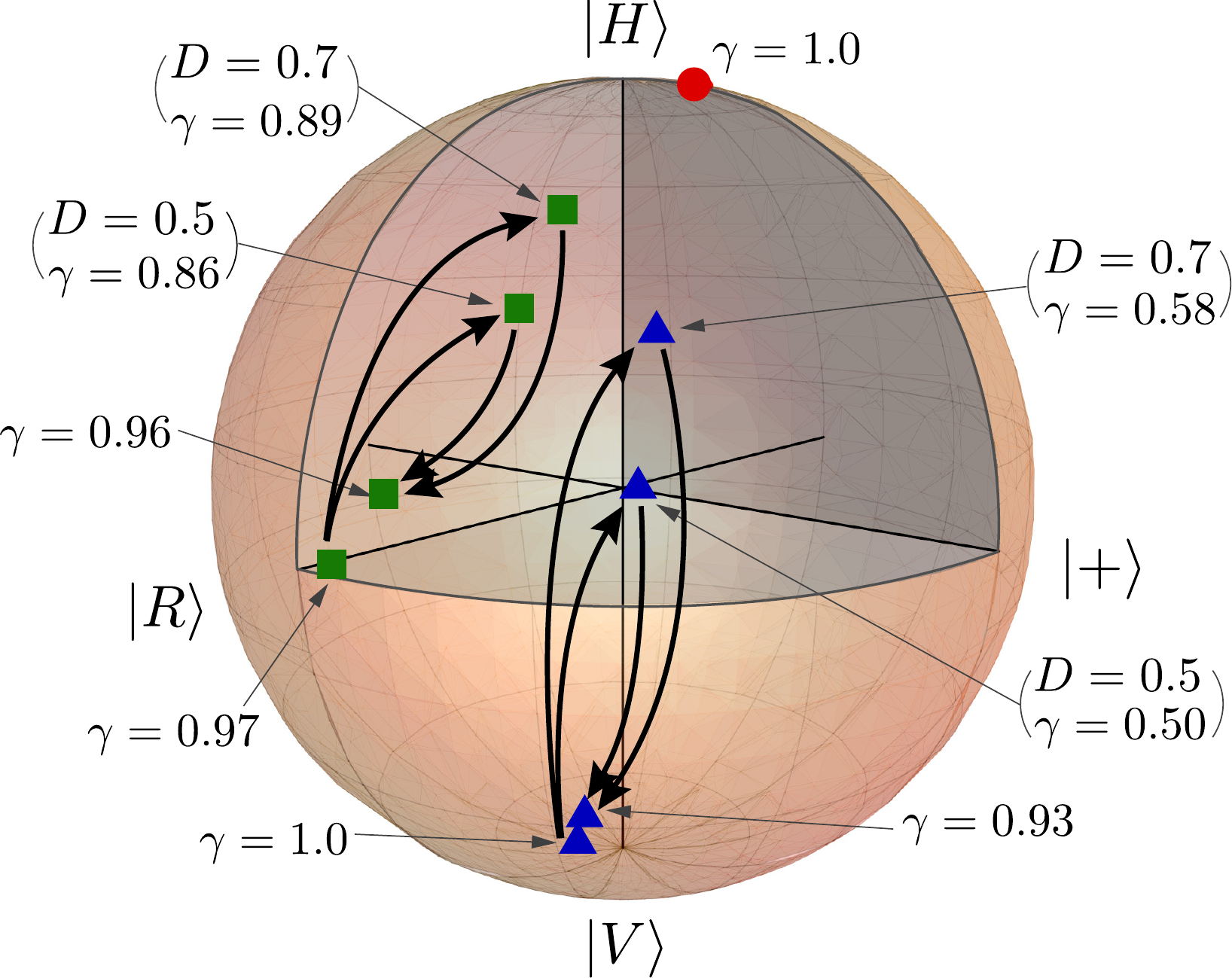}
\caption{Experimental results obtained with QST. The initial qubit states, the states after decoherence $D$, and the states after applying the decoherence suppression scheme are shown. For the decoherence suppression scheme, the weak measurement strength is $p=0.9$ and the reversing measurement strength is optimal $p_r=p+D(1-p)$. In all cases, the fidelity between the input and the recovered states is better than 0.96. The purity of  the state is $\gamma=Tr[(\rho_S^f)^2]$. Note that the initial points do not lie exactly on the poles of the Bloch sphere as they represent experimentally prepared quantum states.}\label{bloch}
\end{figure}

\section{Tomographic analysis}

Using the experimental apparatus described above, we have investigated how the system qubit state $|\psi\rangle_S$ evolves through the decoherence channel and how the decoherence effect can be suppressed via quantum measurement reversal by performing quantum state tomography (QST) and quantum process tomography (QPT).

Fig.~\ref{bloch} shows some of the experimental results obtained with QST. The initial qubit states ($|H\rangle$, $|V\rangle$, and $|R\rangle=(|H\rangle-i|V\rangle)/\sqrt{2}$),  the states after experiencing amplitude damping decoherence with magnitude $D$ for $D=0.5$ and $0.7$, and the states after applying the decoherence suppression scheme (i.e., weak measurement and reversing measurement before and after the decoherence channel, respectively) are shown. For the cases shown here, the weak measurement strength is set at $p=0.9$ and the reversing measurement strength is the optimal value $p_r=p+D(1-p)$. We first point out that the asymmetric nature of amplitude damping decoherence is well demonstrated in Fig.~\ref{bloch}. Under the amplitude damping decoherence, the $|H\rangle$ input state remains unchanged while the $|V\rangle$ and $|R\rangle$ input states are incoherently collapsed towards $|H\rangle$ state. For example, with $D=0.5$, the $|V\rangle$ input state has collapsed to the fully mixed state ($\gamma=Tr[(\rho^f_S)^2]=0.50$). After applying the decoherence suppression scheme, the states are shown to be recovered back to the original states. See the arrows in Fig.~\ref{bloch}. In all cases shown in Fig.~\ref{bloch}, the fidelity, $\mathcal{F}=Tr[\rho_{in}\rho_{rec}]$, between the input state, $\rho_{in}$, and the recovered state, $\rho_{rec}$, is better than 0.96.

\begin{figure}[t]
\centering
\includegraphics[width=4in]{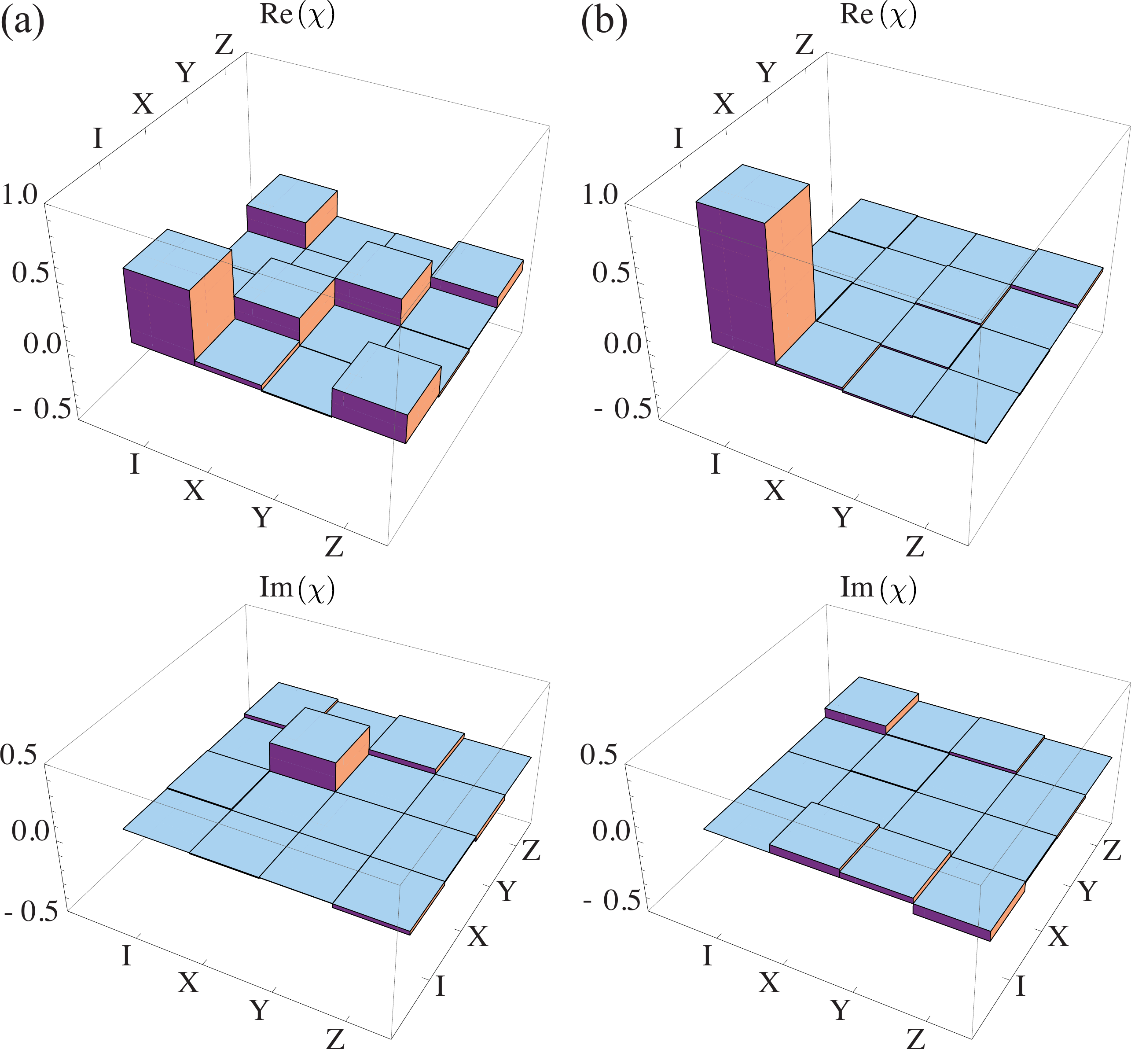}
\caption{The $\chi$-matrices obtained with QPT for (a) the decoherence channel with $D=0.8$ and (b) the decoherence suppressed channel via quantum measurement reversal. For (b), the weak measurement strength is $p=0.9$ and the reversing measurement strength $p_r$ is the optimal value.}\label{chi}
\end{figure}

The quantum processes corresponding to the decoherence channel and the decoherence suppressed channel via quantum measurement reversal can be analyzed by using QPT \cite{Nielsen,Kim}. Since the decoherence suppression operation should ideally retrieve the original quantum state, it corresponds to the identity operation on the input system qubit state. Thus, the performance of the decoherence suppression operation can be quantified by evaluating the process fidelity $F$ between the $\chi$-matrix for the identity operation, $\chi_{I}$, and the experimentally obtained $\chi$-matrix, $\chi_\textrm{exp}$ with the relation $F=Tr[\chi_I \chi_\textrm{exp}]$. For example, Fig.~\ref{chi}(a) shows the $\chi$-matrix of the amplitude damping decoherence channel with $D=0.8$ and, clearly, it is far from the identity operation ($F=0.52$). When the decoherence suppression scheme with the weak measurement strength $p=0.9$ and the optimal value of reversing measurement strength $p_r$ is used, the resulting $\chi$-matrix of the complete process is indeed very close to the identity operation with $F=0.94$, see Fig.~\ref{chi}(b).

\section{Process fidelity and channel transmittance}

\begin{figure}[b]
\centering
\includegraphics[width=3.5in]{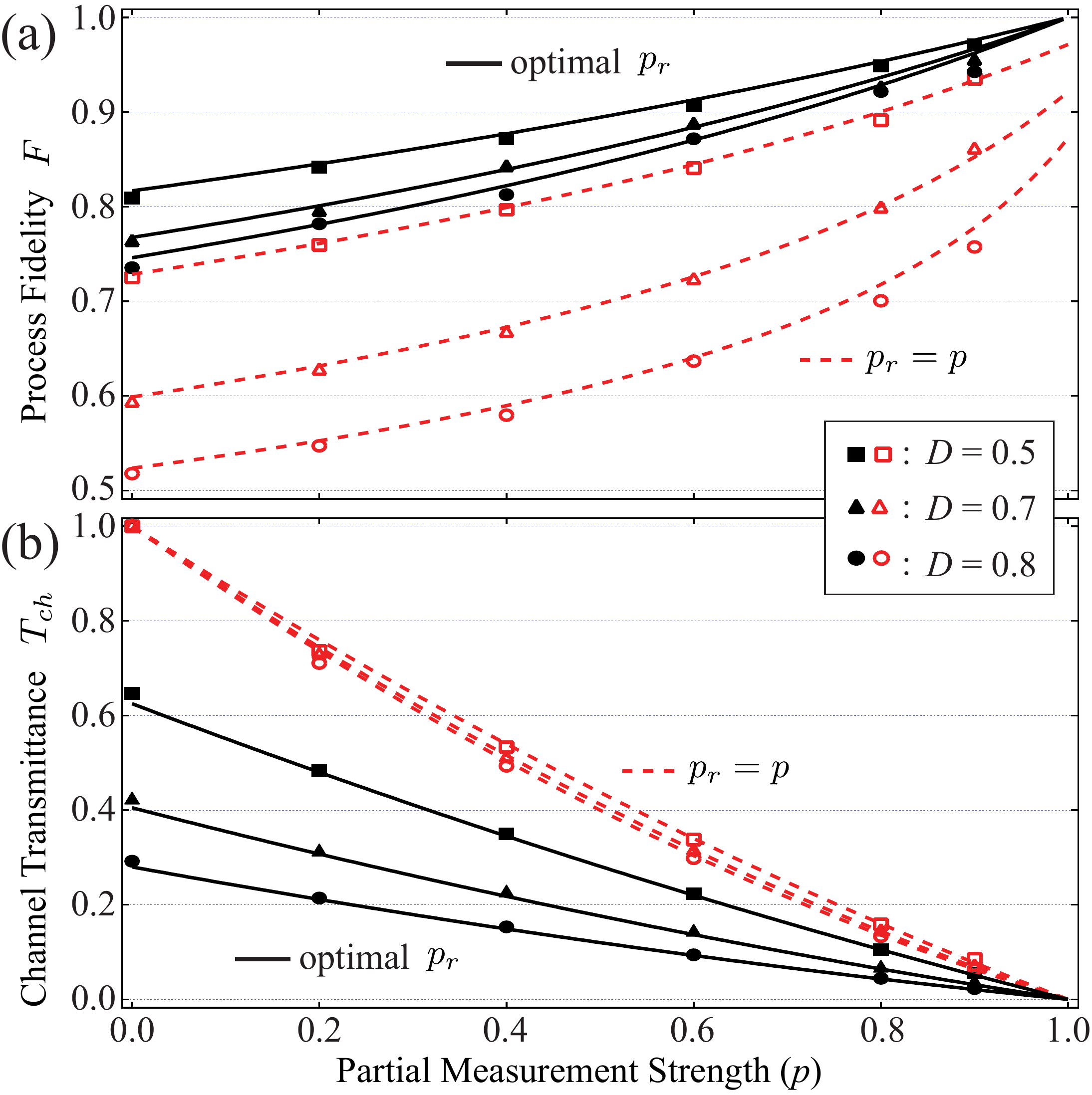}
\caption{(a) The process fidelity of the decoherence suppression scheme. The optimal reversing measurement strength is $p_r=p+D(1-p)$. (b) The corresponding channel transmittance $T_{ch}$. The solid and dotted lines are theoretical curves.}\label{result} 
\end{figure}

The process fidelity $F$, the measure of success of decoherence suppression operation, depends on the weak measurement strength $p$ and the reversing measurement strength $p_r$. Note that, in general, stronger partial collapse to the $|0\rangle_S$ state, i.e., a large $p$ value, would mean better process fidelity. Fig.~\ref{result}(a) shows the theoretical and the experimentally obtained relation between the process fidelity $F$ and the weak measurement strength $p$ for $D=0.5$, 0.7, and 0.8. If the magnitude of decoherence $D$ is not known, the best strategy for the reversing measurement is to choose $p_r=p$. It is, however, possible to perform the reversing measurement that is optimal (in the sense that the resulting process fidelity $F$ is maximal) and this requires \textit{a priori} knowledge of the magnitude of decoherence $D$ as $p_r=p+D(1-p)$. As shown in Fig.~\ref{result}(a), the optimal reversing measurement indeed outperforms the simple reversing measurement strategy with $p_r=p$. The results also show that the stronger the initial weak measurement strength $p$, the better the process fidelity. Again, this is due to the fact that the stronger weak measurement moves the initial system qubit closer to the $|0\rangle_S$ state that does not experience amplitude damping decoherence. For the sufficiently strong partial collapse measurement, $p=0.9$, and with the optimal reversing measurement, very effective decoherence suppression is observed with $F>0.94$, see Fig.~\ref{result}(a).

The non-unitary nature of the weak measurement and the reversing measurement is reflected in the channel transmittance. The channel transmittance is inversely related to the process fidelity. This result is shown in Fig.~\ref{result}(b): The optimal reversing measurement strength $p_r$ results less channel transmittance than choosing $p_r=p$ because the optimal $p_r$ produces higher process fidelity. The trade-off relation between the channel transmittance and the process fidelity is clearly seen in Fig.~\ref{result}.

\section{Conclusion}

We have shown experimentally that amplitude damping decoherence in quantum channels can be effectively suppressed by introducing a weak measurement and the reversing measurement before and after the decoherence channel, respectively. The trade-off relation between the channel transmittance and the process fidelity has also been investigated. 

Although the amplitude damping decoherence and its suppression via quantum measurement reversal were originally discussed in the context of the superconducting qubit \cite{Korotkov}, this type of decoherence is also important for other qubit systems, including photonic qubits and atomic energy qubits. In this paper, we have experimentally verified the validity of the decoherence suppression scheme via quantum measurement reversal using the photonic polarization qubit.  In the case of photonic qubits, however, the decoherence suppression scheme demonstrated here is likely to be more relevant for other types of photonic qubits, such as, the dual-rail qubit (i.e., the path qubit) or the vacuum--single-photon qubit in which probabilistic loss of a photon is tied to amplitude damping decoherence. For the atomic energy qubit, the decoherence suppression scheme appears to be generally applicable since the amplitude damping decoherence is quite natural due to spontaneous emission, similarly to the case of the superconducting qubit. We, thus, believe that the decoherence suppression scheme demonstrated in this work should be widely applicable in quantum information research.

\section*{Acknowledgments}

We would like to thank Dr. Joonwoo Bae for helpful comments. This work was supported, in part, by the National Research Foundation of Korea (2009-0070668 and 2009-0084473), the Ministry of Knowledge and Economy of Korea through the Ultrafast Quantum Beam Facility Program, and the BK21 program.

\end{document}